\documentclass[floatfix,showpacs,showkeys,twocolumn,twoside,prb]{revtex4}
\usepackage{amssymb}
\usepackage{amsmath}
\usepackage{textcomp}
\usepackage{rcs}
\RCS $Date$
\usepackage{comment}
\usepackage[pdftex]{graphicx}

\newcommand{\be}{\begin{eqnarray}} 
\newcommand{\ee}{\end{eqnarray}} 
\newcommand{\bml}{\begin{multline}}
\newcommand{\eml}{\end{multline}}

\newcommand{\bc}{\begin{comment}} 
\newcommand{\ec}{\end{comment}}

\begin{document}

\title{Dielectric response functions of multi-component hot carrier plasmas}
\author{Kyung-Soo Yi}
\email[Corresponding author. e-mail: ]{ksyi@pusan.ac.kr}
\author{Hye-Jung Kim}
\altaffiliation{Present address: Basic Science Research Institute, University of Ulsan, Ulsan 680-749, Korea}
\affiliation{Department of Physics,
Pusan National University, Busan 609-735, Korea}
%\altaffiliation[Corresponding author. e-mail: ]{ksyi@pusan.ac.kr}
%\date{\today}
\date{\RCSDate}

\begin{abstract}
Dielectric responses of a multi-component plasma are investigated at finite temperature considering the case of optically generated carriers in a wurtzite GaN. 
The effective dielectric function and polarizability functions $\Pi_c (q,\omega)$ are determined such that they satisfy the Dyson equations of the effective interactions among plasma components. 
The behavior of plasma component-resolved electronic polarizability functions and the effective dielectric function of the plasma are examined including the effects of dynamic screening at finite temperature. 
Spectral analysis of the dielectric response functions are performed and their dynamic and nonlocal (finite-$q$) behaviors are explored. 
Our result shows that, for a multi-component plasma of high carrier densities, the well-defined optic plasmonic modes are clearly seen well outside the single particle excitation continuum of lighter mass electrons and that the additional low frequency acoustic plasmonic branch is located outside single particle excitation continuum of doubly degenerate heavier mass A-holes. 
The contribution of the heavier mass species to dressed $\Pi_c (q,\omega)$ is very similar to the case of bare $\Pi_c^0 (q,\omega)$.
%For the processes of small frequency exchanges during carrier-carrier scattering, the nonlocal effect of the screening is negligible reducing the case to that of Thomas-Fermi screening limit.

\end{abstract}
\maketitle

%\def\thefootnote{\fnsymbol{footnote}}
%\setcounter{footnote}{0}
%\footnotetext[1]{Corresponding author. e-mail: jskim99@pusan.ac.kr}
\vspace{-0.5cm}
\section{Introduction}
With the use of high power ultra short pulse lasers, degenerate electron--hole plasmas could be readily generated in semiconductor materials.\cite{Kyhm2011}
%, and problems of hot carriers in semiconductors have been investigated much in detail, especially in GaAs material. Recently, physics of photo-generated hot carriers in wide band gap semiconductors such as ZnO and GaN have gained renewed interest because of their practical importance in application to short wavelength optoelectronic devices \cite{Hagele,Ozgur,Sarua}.
In these plasmas, the carriers are highly excited to occupy the conduction and valence bands forming multi-component carrier system separated in reciprocal space. 
%The electronic Coulomb interaction renormalizes the bare interactions among electrons giving rise to screened interactions.
Transport properties and energy relaxation of the carriers strongly depend on the nature of the carrier screening,\cite{Hwang2007,Wunsch} which affects the electrical polarizability and dielectric function of the multi-component plasma. 
%However, important issues on major basic properties, such as dielectric responses of the multi-component hot carriers in these materials are not well detailed yet to provide information about the carriers in such wide band gap material.

In the presence of two or three different types of plasma components, the dielectric screening and individual energy dissipation behavior become so involved (compared to the case of single-component carriers) to be included properly in the model.\cite{Prunnila2007}
In an electron-hole plasma, in addition to the well known optic plasmon-LO phonon coupling of the single-component plasma such as in doped semiconductors,\cite{DasSarma1990} low frequency acoustic collective oscillations also occur.\cite{Platzman1973,Abstreiter1984}
Here, we limit our analysis on the finite-temperature dielectric responses, which are of importance in the optical and transport responses of many carrier systems, 
% applied to the case of quasi-thermalized hot electrons and holes in wurtzite GaN 
focusing %our attention 
to the effects of many-body dynamic carrier screening on the component-resolved electronic polarizability functions and effective dielectric function of the plasma.
\begin{figure*}[t]
\includegraphics[width=15cm]{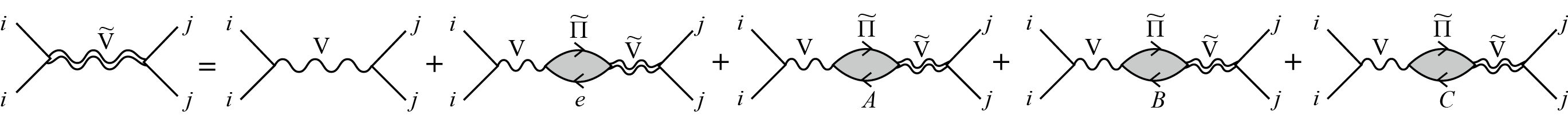}
\caption{Dyson equations for the effective (dressed) interactions $\tilde{V}_{ij}$ between carriers of types $i$ and $j$, where carrier indices $i$ and $j$ can be either electron, valence band holes in the A-, B-, or C-band of the material.}
\label{dyson}
\end{figure*}
%
%
\begin{comment}
[[
The results are used to study the finite-temperature spectral behavior of phonon spectral functions of the multi-component hot carriers.  
The screening of the electron--phonon interactions is included by assuming that the carriers respond as independent particles to the total effective electrostatic potential. 
Carrier-phonon coupling channels of polar and nonpolar optical phonons, acoustic deformation-potential and piezoelectrical Coulomb couplings to the excited electrons and holes are considered. 
%In this paper we concentrate on the influence of dynamic screening and plasmon-phonon mode coupling in the finite-temperature dielectric responses and phonon spectral functions of a multi-component hot carrier plasma. 
]]
\end{comment}
%In Sec. II, we present calculation methods and discuss results for the case of optically excited wurtzite GaN, and we sum up the results in Sec. III. 

%  

%\section{Formulation}
\section{Dielectric response functions}
\begin{comment} In a multicomponent plasma such as in photo-excited plasmas in intrinsic semiconductors, the low frequency acoustic mode of collective oscillations occurs,\cite{Platzman1973,Abstreiter1984} in addition to the optic plasmon-LO phonon coupling of the single-component plasma such as of doped semiconductors.\cite{DasSarma1990}
\end{comment}
We consider the case that hot electrons and holes are generated in the conduction and valence bands, respectively, by optically pumping the material.  
The carrier-carrier interaction introduces dielectric screening and, hence, affects various carrier-phonon coupling channels of the material.
In a multi-component many carrier system (consisting electrons and various holes), we note that the effective (dressed) interactions $\tilde{V}_{ij}$ between carriers of types $i$ and $j$ are the solutions of the Dyson equation,\cite{Fetter}
\be
\tilde{V}_{ij}=V_{ij}+ \sum_{\ell=\rm e, A, B, C} V_{i\ell}\tilde{\Pi}_{\ell\ell}\tilde{V}_{\ell j};\mbox{  } \tilde{V}_{ij}=\frac{V_{ij}}{\tilde\epsilon},
\label{V dyson}
\ee
where $V_{ij}$ is the bare (lowest order) Coulomb interaction $v_q (\equiv \frac{4\pi e^2}{q^2})$ for $i=j$ or $-v_q$ for $i\neq j$, and $\tilde{V}_{ij}$ includes the polarization effects of the medium.
Here carrier indices $i$ and $j$ can be either electrons (denoted by `e') or holes in the A, B, or C band of the material.
The Dyson equation for dressed Coulomb interactions $\tilde{V}_{ij}$ between conduction electrons and various holes are illustrated in Fig. \ref{dyson} in terms of (retarded) proper polarization functions $\tilde{\Pi}_{\ell\ell}$ (or density--density correlation function $\hbar\tilde{\Pi}_{\ell\ell}$)  and bare Coulomb interactions $V_{ij}$ among plasma components.

One can solve the coupled equations represented by Eq.(\ref{V dyson}) for $\tilde{V}_{ij}$ to obtain the dielectric function $\tilde\epsilon$ of the multi-component many carrier system written as
\be
\tilde\epsilon (\vec q,\omega) = 1-v_q \sum_{\nu=\rm e, A, B, C} \tilde{\Pi}_{\nu\nu}(\vec q,\omega).
\label{epsilon}
\ee
Here we note that the proper polarization function $\tilde{\Pi}_{\nu\nu}$ of each plasma component (denoted by $\nu$) contributes to the effective dielectric function of the (multi-component) plasma.
\begin{comment}
\begin{align}
\Pi_{\ell\ell}(\vec q,\omega)= & \frac{\Pi_{\ell\ell}^0(\vec q,\omega)}{\tilde\epsilon(\vec q,\omega)},
\label{polarization}
\end{align}
\end{comment}

The (retarded) full polarization function $\Pi_{\nu\nu'}(\vec q,\omega)$ is the Fourier transform of the (retarded) density-density (time) correlation function defined by\cite{Fetter}
\be
\hbar\Pi_{\nu\nu'}(\vec q,t)
=-i \theta(t)&\frac{1}{\mathcal{V}} \left< [\hat n_{\vec q,\nu}(t),\hat n_{-\vec q,\nu'}(0)]\right>, 
\label{time correlation theorem}
\ee
where $\theta(t)$, $\mathcal{V}$, and $\hat n_{q,\nu}(t)$ are the Heaviside unit step function, volume of the sample, and the number operator in the Heisenberg representation of the carriers in the band $\nu$.
The full polarization $\Pi_{\nu\nu'}(\vec q,\omega)$ satisfies Dyson equation given, in terms of proper  counterpart $\tilde{\Pi}_{\nu\nu'}(\vec q,\omega)$, by\cite{Bruus}  
\be
\Pi_{ij}=\tilde{\Pi}_{ij} + \sum_{\ell m}\tilde{\Pi}_{i\ell} V_{\ell m}\Pi_{m j}, %;\mbox{  } \Pi_{ij}=\frac{\Pi_{ij}^{irr}}{\tilde\epsilon},\
\label{Pi dyson}
\ee
implying that the full polarization $\Pi_{\nu\nu'}$ need be determined self-consistently. 

The real part of the electric conductivity, $\mathcal{R}e \sigma_{\nu\nu'} (\vec q,\omega;T)$, a measure of dissipative processes, is directly linked to the imaginary part of the retarded polarization function, $\mathcal{I}m \Pi_{\nu\nu'} (\vec q,\omega;T)$ %, i.e., the spectral function $\mathcal{A}(q,\omega)=-2\mathcal{I}m\Pi_{\nu\nu}(\vec q,\omega)$ 
of the plasma, in which quanta of wavenumber $\vec q$ and frequency $\omega$ are absorbed by the carriers in the plasma.
%we note that or the retarded polarization function $\Pi_{\nu\nu}(\vec q,\omega)$ satisfies the fluctuation--dissipation theorem written as\cite{Zubarev,Doniach}
Once the range of transparency windows for the energy dissipation is known, it can be utilized to engineer the waveguide characteristics of the material.
In this work, $\tilde{\Pi}$ denotes temperature-dependent proper (`irreducible' in the language of Ref.\cite{Bruus}) polarization function of the multi-component many carrier system.

In the random phase approximation (RPA), which is supposed to be valid in the limit of high carrier densities, $\tilde{\Pi}_{\nu\nu'}$ is approximated by the bare (noninteracting) counterpart $\Pi_{\nu\nu'}^0$, which is given, if evaluated at a carrier temperature $T_{\rm ch}$, by
\cite{Fetter,Giuliani}%[{Fetter Eq15-9,p181},Mahan 3rd 5-152 p328;Giuliani4-16,p160]
\be
\Pi_{\nu\nu'}^{0}(\vec q,\omega) = -\frac{2}{\hbar}\sum_{\vec k}\frac{f_{k+q,\nu}^{(0)}-f_{k,\nu'}^{(0)}}{\omega-(\omega_{k+q,\nu}^{(0)}-\omega_{k,\nu'}^{(0)})+i\eta}. 
\label{Pi0}
\ee
The Lindhard-type expression of Eq.(\ref{Pi0}) is analogous to the case of spin-resolved expression in a multi-component spin system.\cite{Yi-Quinn1996}
The RPA to the dielectric function is written as \cite{Collet1993} 
\be
\epsilon_{\rm RPA} (\vec q,\omega) = 1-v_q \sum_{\nu=\rm e, A, B, C} \Pi_{\nu\nu}^0(\vec q,\omega). 
\label{rpa epsilon} 
\ee  
The explicit expression of the lowest order proper polarization function $\Pi_{\nu\nu'}^{0}(\vec q,\omega)$ can be written in terms of its zero-temperature Lindhard function\cite{Maldague}, 
and the real and imaginary parts are given by\cite{Giuliani},
\begin{align}
\mathcal{R}e \Pi_{\ell\ell}^0 (\vec q,\omega;T) =& -g_{\ell} \int_0^\infty dx \frac{F(x,T)}{q/k_{F\ell}}  \nonumber\\
\times 
\frac{1}{2}&\left[\ln \left|\frac{x-\nu_{\ell-}}{x+\nu_{\ell-}}\right|-\ln \left|\frac{x-\nu_{\ell+}}{x+\nu_{\ell+}}\right| \right]
\label{Re Pi0}
\end{align}
and
\begin{align}
\mathcal{I}m \Pi_{\ell\ell}^0 &(\vec q,\omega;T) \nonumber \\
=-\pi g_{\ell} &\left[\frac{\omega}{v_{F\ell}  q}
+\frac{k_B T}{\hbar v_{F\ell}  q}
\ln \frac{1+e^{\beta[\nu_{\ell-}^2\varepsilon_{F\ell}  - \mu_{\ell} (T)]}}{1+e^{\beta[\nu_{\ell+}^2
\varepsilon_{F\ell}  - \mu_{\ell}(T)]}}\right],
\label{Im Pi0}
\end{align} 
where $F(x,T)=\frac{x}{e^{\beta[x^2\varepsilon_{F\ell} - \mu_{\ell}(T)]}+1}$, 
$g_{\ell}=\frac{m_{\ell}k_{F\ell}}{\pi^2\hbar^2}$, $k_{F\ell}=(3\pi^2 n_\ell)^{1/3}$,
and $\nu_{\ell\pm}=\frac{\omega}{qv_{F\ell}} \pm\frac{q}{2k_{F\ell}}$.
In Eqs. (\ref{Re Pi0}) and (\ref{Im Pi0}), conventional notations are used such as $m_{\ell}$, $k_{F\ell}$, $v_{F\ell}$, $\varepsilon_{F\ell}$, and $\mu_{\ell}$ denoting effective mass, Fermi wave number, Fermi velocity, Fermi energy, and chemical potential, respectively, of a carrier type indicated by subscript $\ell$ (either electrons or holes) and $T$ is the temperature of the corresponding carriers in quasi-equilibrium. 

\begin{figure*}[t]
\includegraphics[width=7cm]{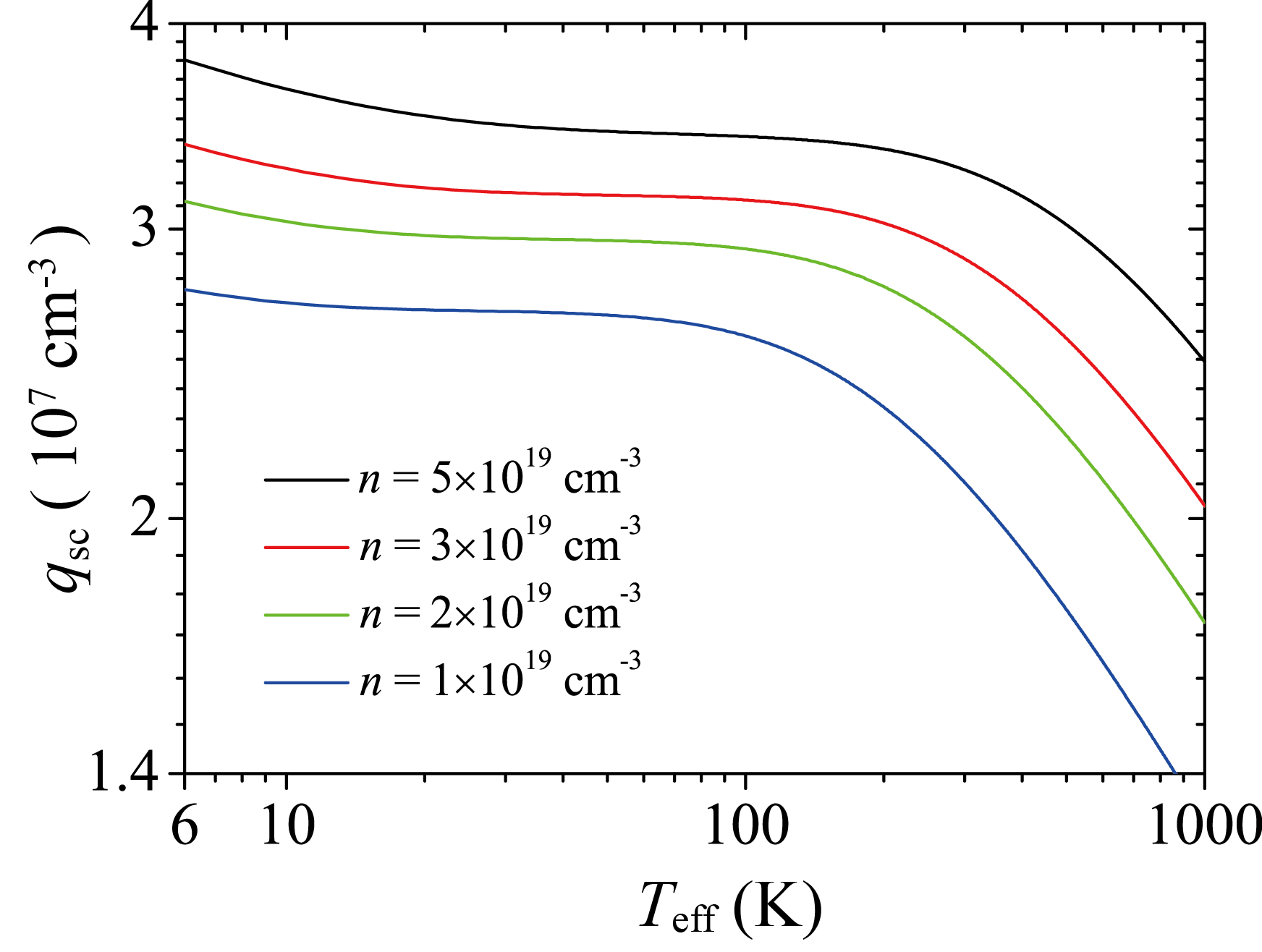}
\caption{Temperature dependence of the Thomas-Fermi screening wavelength $q_{\rm sc}$ of a wurtzite GaN with four different conduction electron densities of $1-5 \times 10^{19} \rm cm^{-3}$.}
\label{qsc}
\end{figure*}
\begin{figure*}[t]
\includegraphics[width=7cm]{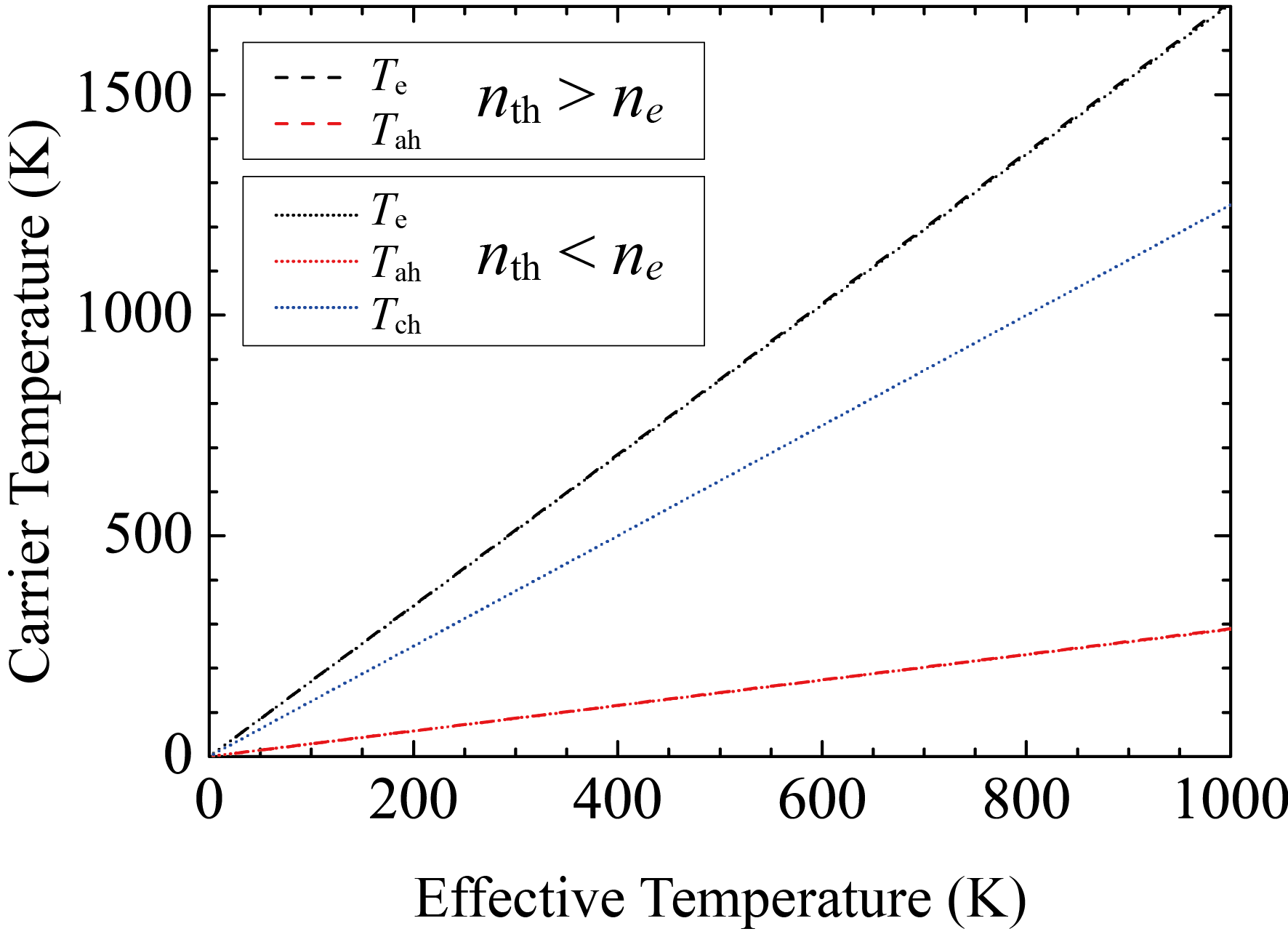}
\caption{Relation between carrier temperatures ($T_e, T_{ah}, {~\rm and~} T_{ch}$) and the effective temperature $T_{\rm eff}$ of the multi-component plasma in a wurtzite GaN for $n_{e}= 2 \times 10^{19} \rm cm^{-3}$.}
\label{Teff}
\end{figure*}
For the case of a single-component plasma (scp), Eq.(\ref{rpa epsilon}) reduces to 
\be
\epsilon_{\rm RPA}^{(\rm scp)} (\vec q,\omega) = 1-v_q \Pi^0(\vec q,\omega), 
\label{epsilon single}
\ee
and we have 
$
\mathcal{R}e \,\Pi_{\rm RPA}^{(\rm scp)} = \frac{\mathcal{R}e \,\Pi^{0}-v_q |\Pi^{0}|^2}{|\epsilon_{\rm RPA}^{(\rm scp)}|^2}
$
and
$\mathcal{I}m \,\Pi_{\rm RPA}^{(\rm scp)} = \frac{\mathcal{I}m \,\Pi^{0}}{|\epsilon_{\rm RPA}^{(\rm scp)}|^2}$.
If we take the static approximation for screening, $\epsilon_{\rm RPA}(\vec q,0)$ is given, in the long wave length limit, 
$
\epsilon_{\rm RPA}(\vec q,0)=1+\frac{q_{\rm sc}^2}{q^2}.
$ 
Here $q_{\rm sc}$ is the temperature-dependent Thomas-Fermi screening wavelength given by
$
q_{\rm sc}^2 = \frac{4\pi e^2}{\epsilon_{\infty}}\sum_{\nu}\frac{\partial n_\nu}{\partial \mu_\nu},
\label{qDH}
$
where $n_\nu$ and $\mu_\nu$ are, respectively, the particle number density and quasi chemical potential of the band $\nu$ occupied by electrons and holes. 
The $q_{\rm sc}$ is a decreasing function of temperature and $q_{\rm sc} \simeq 3.07 k_F$ at 300 K with an electron concentration of $n_{\rm c} =2\times 10^{19} \rm cm^{-3}$.  (See Fig. \ref{qsc}.)
\begin{figure*}[ht]
\includegraphics[width=14cm]{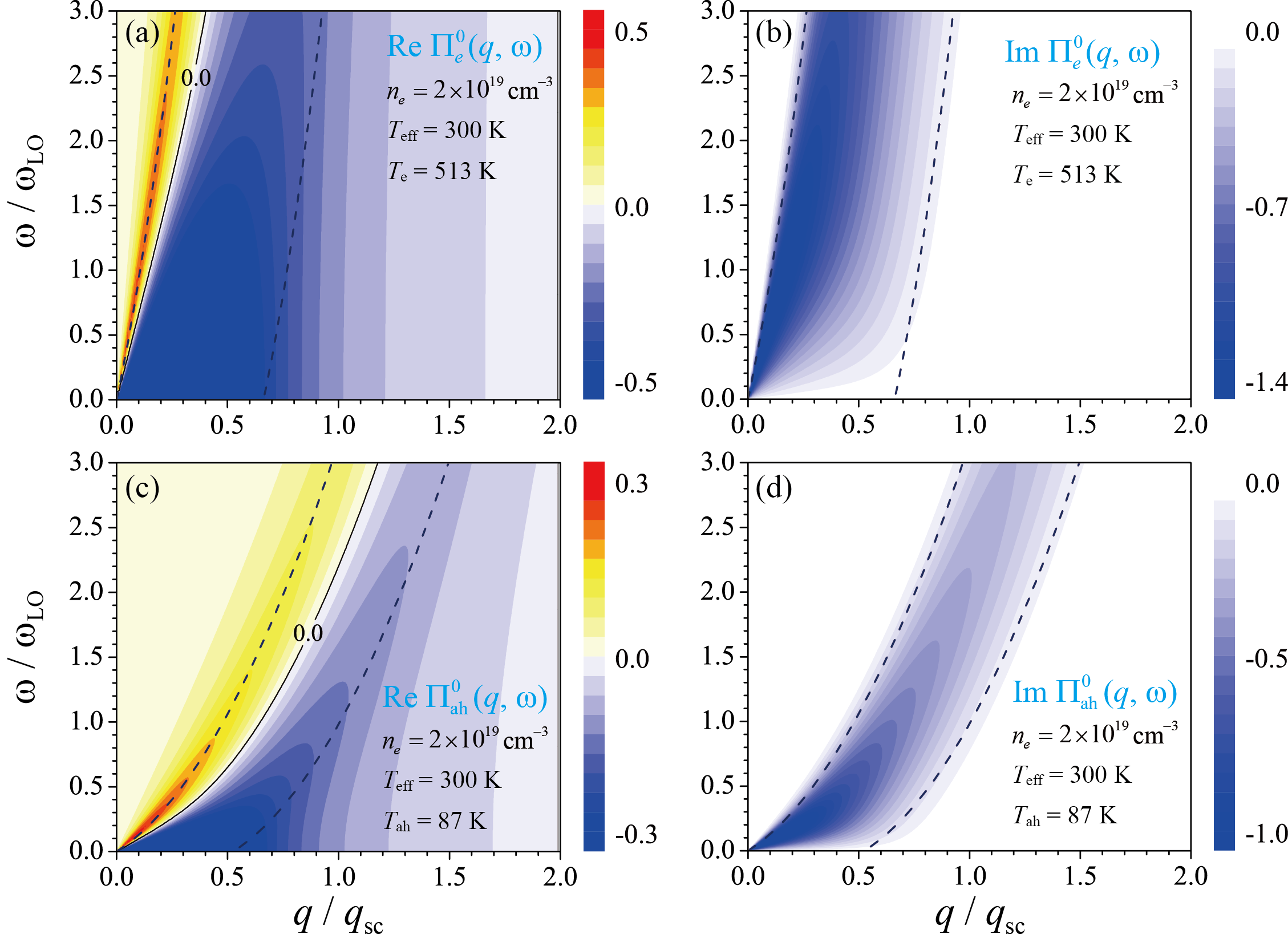}
\caption{Real and imaginary parts of noninteracting single-component plasma polarization functions $\Pi_c^0 (q,\omega)$ for the carrier density of $2 \times 10^{19} \rm cm^{-3}$ at carrier effective temperature $T_{\rm eff}$=300 K.
(a) and (c): the case of conduction electrons (c) and (d): the case of A-holes. Pair of dashed lines denotes the region of allowed corresponding single particle excitations in a wurtzite GaN.}
\label{P0_ri_wq_300K}
\end{figure*}
\begin{figure*}[ht]
\includegraphics[width=14cm]{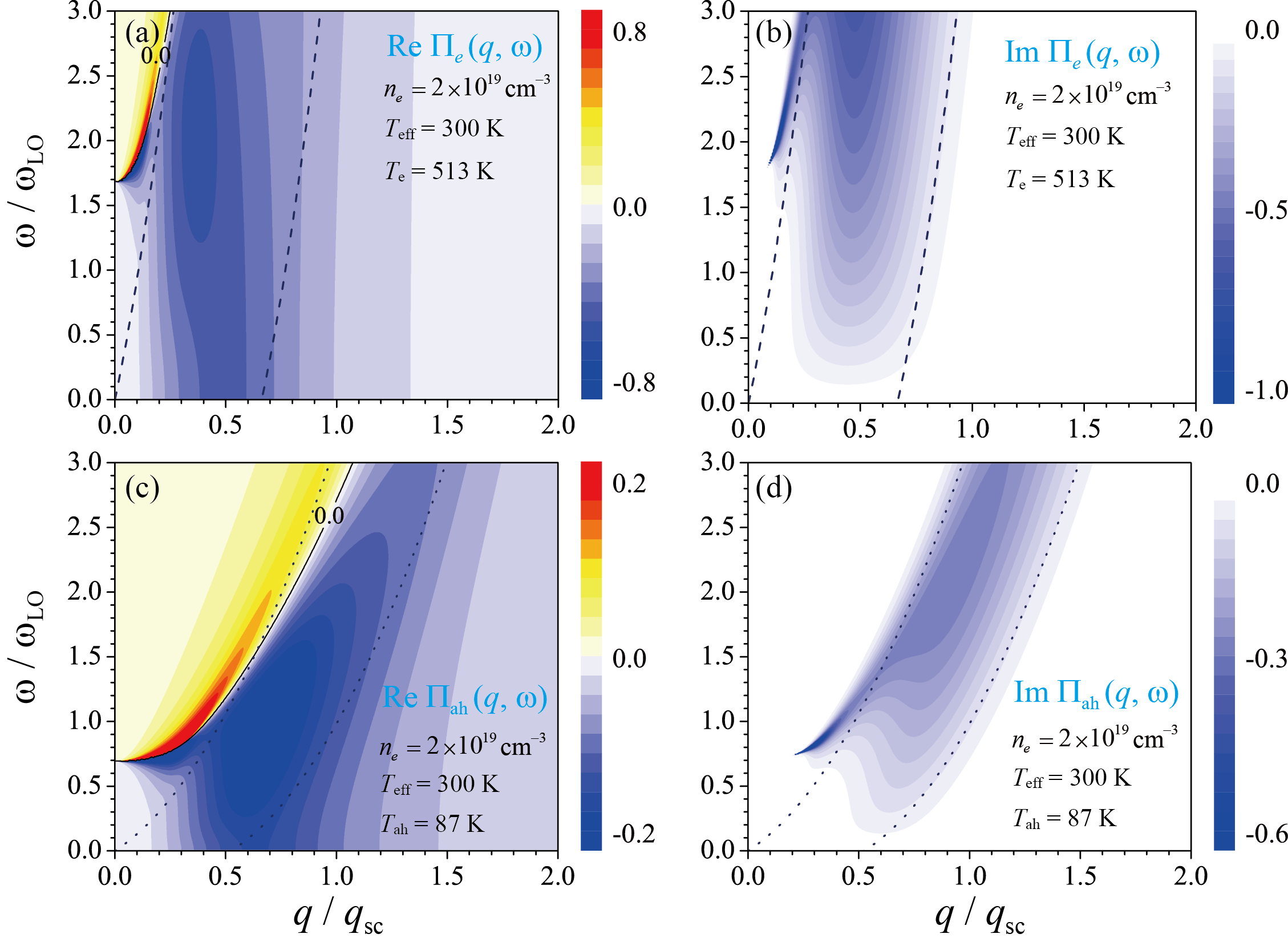}
\caption{Real and imaginary parts of dressed single-component plasma polarization functions $\Pi_c^{\rm scp} (q,\omega)$ for the carrier density of $2 \times 10^{19} \rm cm^{-3}$ at carrier effective temperature $T_{\rm eff}$=300 K. (a) and (b): the case of conduction electrons, (c) and (d): the case of A-holes. Pair of dashed lines denotes the region of allowed corresponding single particle excitations for carriers in a wurtzite GaN.}
\label{PI_scp_ri_wq_300K}
\end{figure*}
\section{Carrier densities and effective temperature of a multi-component plasma}
Here we consider the case of an ideal GaN in the wurtzite phase,\cite{De2010,suzuki1995,chuang1996} where the spin-orbit coupling is of an order of magnitude smaller than the crystal field splitting. Then, the higher lying valence (A and B hole) bands are doubly degenerated in energy under the influence of the hexagonal crystal field and the top of the C hole band is located below by $\Delta$ from the top of the degenerate (A and B hole) bands. 
The carrier concentration in each band now satisfies the relation 
$n_{\rm e} = 2 p_{\rm A} + p_{\rm C} \Theta(n_{\rm e} - n_{\rm th})$.
Here $n_{\rm e}$, $p_{\rm A}(\equiv p_{\rm B})$, and $p_{\rm C}$ are the carrier concentrations in the conduction and valence bands, each given by
\begin{align}
p_{\rm A(B)} =\int_0^{\Delta+\varepsilon_{\rm F A}} dE \,g_{\rm A}(E) %
=\frac{n_{\rm e}+\frac{n_{\rm th}}{2}\left(m_{\rm C}/m_{\rm A}\right)^{3/2}}{2+\left(m_{\rm C}/m_{\rm A}\right)^{3/2}},
\label{nA}
\end{align}        
\begin{align}
p_{\rm C} =\int_\Delta^{\Delta+\varepsilon_{\rm F C}} dE \,g_{\rm C}(E) % \nonumber\\
=\frac{(n_{\rm e}-n_{\rm th})\Theta(n_{\rm e} - n_{\rm th})}{1+2\left(m_{\rm A}/m_{\rm C}\right)^{3/2}},
\label{nC}
\end{align}         
where $g_{\rm A}$($g_{\rm C}$) denotes the density of states for band A(C), and 
\begin{align}
n_{\rm th} =\frac{2}{3\pi^2}\left(\frac{2 m_{\rm A} \Delta}{\hbar^2}\right)^{3/2}.
\label{nth}
\end{align}          
In Eqs. (\ref{nA}) and (\ref{nC}), $n_{\rm th}$ denotes the electron concentration, at which C holes need to appear in the C hole band at higher levels of the optical pumping power.      
              
Since carrier temperatures for different types of carriers are not identical, in general, in a multi-component carrier system, the carrier effective temperature $T_{\rm eff}$ of the sample would be meaningful in experiments\cite{Collet1989}, where $T_{\rm eff}$ is a measure of the average kinetic energy per carrier in the sample.  
In terms of quasi-equilibrium carrier temperatures $T_{\rm e}$, $T_{\rm ah}(=T_{\rm bh})$, and $T_{\rm ch}$, the carrier effective temperature $T_{\rm eff}$ of the multi-component carriers is written as
\begin{align}
T_{\rm eff}=\frac{T_{\rm e}}{2}\left[1+\frac{m_{\rm e}}{n_{\rm e}}\left(2\frac{p_{\rm A}}{m_{\rm A}}+\frac{p_{\rm C}}{m_{\rm C}}\right)\right], 
\label{T_eff}
\end{align}          
where we have noted the energy balance relation of $\hbar\omega_{\rm ph}=\varepsilon_{\rm g}+\Delta E_{\rm e} +\Delta E_{\rm h}$ and $\Delta E_{\rm e}/\Delta E_{\rm h} =m_{\rm h}/m_{\rm e}=T_{\rm e}/T_{\rm h}$.
Here  $\hbar\omega_{\rm ph}$, $\varepsilon_{\rm g}$, $\Delta E_{\rm e}$, and $\Delta E_{\rm h}$ indicate, respectively, the photon energy of the pumping laser, band gap of the material, kinetic energies of the photo-generated electron and the corresponding hole (h=A{\,\rm or\,} C in our case).    
In a wurtzite GaN, we estimate $n_{\rm th} \sim 4.4\times 10^{19} \rm cm^{-3}$.
For the case $n_{\rm e} < n_{\rm th}$, $n_{\rm e} = 2 p_{\rm A}$ to write $T_{\rm eff}=\frac{T_e}{2}\left(1+\frac{m_e}{m_{\rm A}}\right) \approx \frac{7}{12}T_{\rm e}\simeq 3.5T_{\rm ah}$ with $m_{\rm A} \simeq 6 m_{\rm e}$. 
Figure \ref{Teff} illustrates the relation between quasi-equilibrium carrier temperatures $T_{\rm e}$ of the conduction electrons, $T_{\rm A(B)}$ of the A(B) holes, and $T_{\rm ch}$ of the C holes as a function of the effective temperature $T_{\rm eff}$ of the multi-component carriers in a wurtzite GaN at $n_{e}= 2 \times 10^{19} \rm cm^{-3}$.
In this work, the C hole band is taken to be occupied by the carriers when $n_{\rm e}$ is larger than the threshold density $n_{\rm th}$, so that $T_{\rm ch}$ is finite only for $n_{\rm e} > n_{\rm th}$. 

\section{Results and Discussion}
In illustrating our numerical results of the response functions for a wurtzite GaN, we have used the effective masses of $m_{\rm e} =0.22 m_0$, $m_{\rm A(B)}=1.30 m_0$, and $m_{\rm C=}0.3 m_0$ for electrons, doubly degenerate A(B)-holes, and C-holes, respectively,\cite{Landolt,Kyhm2011} where $m_0$ is the mass of a free electron.
In GaN, the formation of hot electron-hole multi-component plasma is expected for carrier densities larger than the Mott density of $10^{18}-10^{19} \rm cm^{-3}$.\cite{Binet1999}

In Fig. \ref{P0_ri_wq_300K}, real and imaginary parts of the noninteracting polarization function $\Pi_c^0 (q,\omega)$ defined by Eq.(\ref{Pi0}) for electrons (e) and A-holes (ah) are illustrated for conduction electrons of $2 \times 10^{19} \rm cm^{-3}$ at carrier effective temperature $T_{\rm eff}$ of 300 K.
The wave number and frequency are scaled by the Thomas-Fermi screening wave number $q_{\rm sc}$ (cf. Fig. \ref{qsc}) and the longitudinal bare phonon frequency $\omega_{\rm LO}$, respectively.
Boundaries of allowed single particle excitations for electrons or holes are designated by a pair of dashed lines in each figure.
Within the continuum region, electrons (holes) within the  Fermi sea can be excited to states outside the Fermi sea. 
The zero of $\mathcal{R}e \Pi_c^0 (q,\omega)$ is indicated by a solid line within the continuum region of the single particle excitations, and, at small $q$, $\mathcal{R}e \Pi_c^0 (q,\omega)$ changes sign from negative to positive as $\omega$ increases sweeping across the continuum region. (See Fig. \ref{P0_ri_wq_300K}(a) for $\mathcal{R}e \Pi_e^0 (q,\omega)$ and (b) for $\mathcal{R}e \Pi_{ah}^0 (q,\omega)$.) 
On the other hand, $\mathcal{I}m \Pi_c^0 (q,\omega)$ is significantly different from zero and negative in the continuum region for the single particle excitations showing peak structure along the zero line of  $\mathcal{R}e \Pi_c^0 (q,\omega)$ indicated in panels (a) and (c). 
(See Fig. \ref{P0_ri_wq_300K} (b) for $\mathcal{I}m \Pi_e^0 (q,\omega)$  and (d) for $\mathcal{I}m \Pi_{ah}^0 (q,\omega)$.)  
This observation is a direct consequence of the Kramers-Kronig dispersion relations.\cite{Giuliani} 
In the absence of carrier screening, the result of $\Pi_c^0 (q,\omega)$ shows clearly that single particle excitations of free electrons or free holes are the only process for the energy and momentum dissipation, because the effects of carrier screening is completely ignored in $\Pi_c^0 (q,\omega)$. 
\begin{figure*}[ht]
\includegraphics[width=14cm]{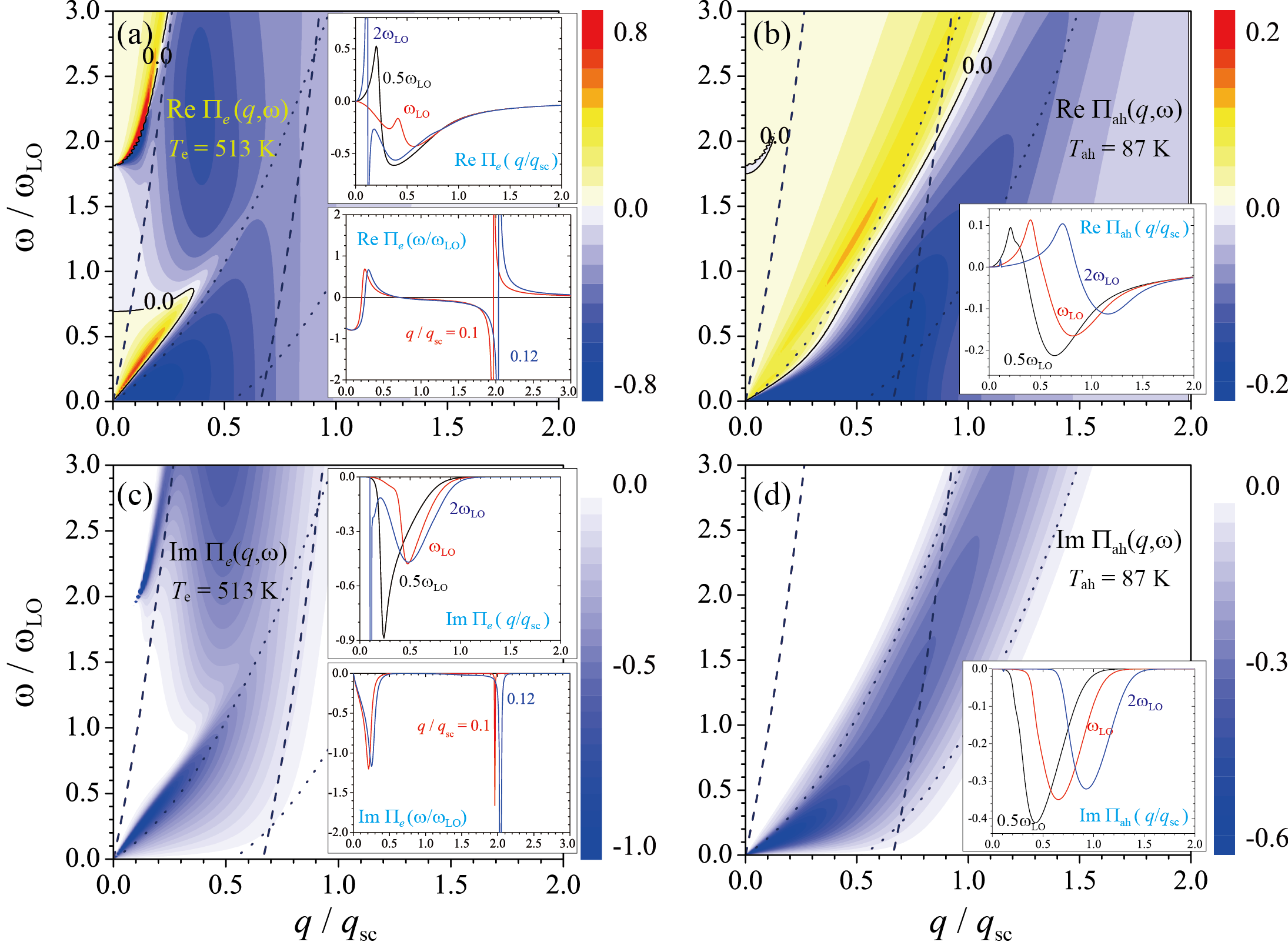}
\caption{Real and imaginary parts of dressed polarization functions $\Pi_c (q,\omega)$ for conduction electrons [panels (a) and (c)] and A-holes [panels (b) and (d)] at electron concentration of $2 \times 10^{19} \rm cm^{-3}$ and carrier effective temperature 300 K. Insets illustrate cross-sectional view of the spectral behavior at $q=0.1$ and 0.19 $q_{\rm sc}$, where $q_{\rm sc} = 2.58 \times 10^{7} \rm cm^{-1}$ at 300 K.   
Pairs of dashed and dotted lines denote the allowed regions of single particle excitations for electrons and A-holes, respectively, in a wurtzite GaN.}
\label{PI_ri_wq_300K}
\end{figure*}

Combining Eqs.(\ref{Pi dyson}) and (\ref{Pi0}), one can write the retarded full polarization, in the RPA, for a carrier of type $\ell$, since $\tilde \Pi_{ij} =\Pi_{ij}^0 \delta_{ij}$ in the RPA, as\cite{Collet1989}
\be
\Pi_{\ell\ell}=\Pi_{\ell\ell}^0 \left[1 + \frac{v_q \Pi_{\ell\ell}^0}{1-v_q\sum_{\nu=\rm e, A, B, C} \Pi_{\nu\nu}^0}\right].
\label{full polarization}
\ee
We note that, in a multi-component plasma, each plasma component (indicated by the index $\nu$ in the denominator) contributes to the polarization $\Pi_{\ell\ell}$ of a given carrier component denoted by $\ell$.
The denominator on the right hand side of the above expression is the dielectric function of the plasma in the RPA defined in Eq.(\ref{rpa epsilon}).  
In the RPA, one can express $\mathcal{R}e \Pi_{\ell\ell} (\vec q,\omega;T)$ and $\mathcal{I}m \Pi_{\ell\ell} (\vec q,\omega;T)$ in terms of $\mathcal{R}e \Pi^0_{\nu\nu}$ of Eq.(\ref{Re Pi0}) and $\mathcal{I}m \Pi^0_{\nu\nu}$ of Eq.(\ref{Im Pi0}), and they are connected to each other by the Kramers--Kronig relation, the latter (former) being the absorptive part related directly, for example, to the spectral damping behavior (dispersion relation) of collective oscillations of the system.
In a single-component plasma, the dressed polarization function Eq.(\ref{full polarization}) reduces to 
\[
\Pi_{\ell\ell}^{\rm (scp)}={\Pi_{\ell\ell}^0}/(1-v_q\Pi_{\nu\nu}^0) \equiv{\Pi_{\ell\ell}^0}/{\epsilon_{\rm RPA}^{\rm (scp)}},
\]
where $\epsilon_{\rm RPA}^{\rm (scp)}$ is defined by Eq.(\ref{epsilon single}). 
Figure \ref{PI_scp_ri_wq_300K} illustrates the real and imaginary parts of dressed single-component plasma polarization functions $\Pi_c^{\rm scp} (q,\omega)$ at $T_{\rm eff}$=300 K in a wurtzite GaN for the carrier density of $2 \times 10^{19} \rm cm^{-3}$. 
The cases for conduction electrons plasma are shown in panels (a) and (b), and the cases of A-holes in (c) and (d), where 
the contributions from the conventional optic plasmon-LO phonon coupling as in doped semiconductors are shown.\cite{DasSarma1990}
We observe that dynamic screening introduces drastic changes to the bare polarization functions (shown in Fig. \ref{P0_ri_wq_300K}) giving rise to additional energy dissipation channel of (optical) plasmonic collective excitations along with that of single particle excitation continuum indicated by a pair of dashed lines.

\begin{figure*}[t]
\includegraphics[width=14cm]{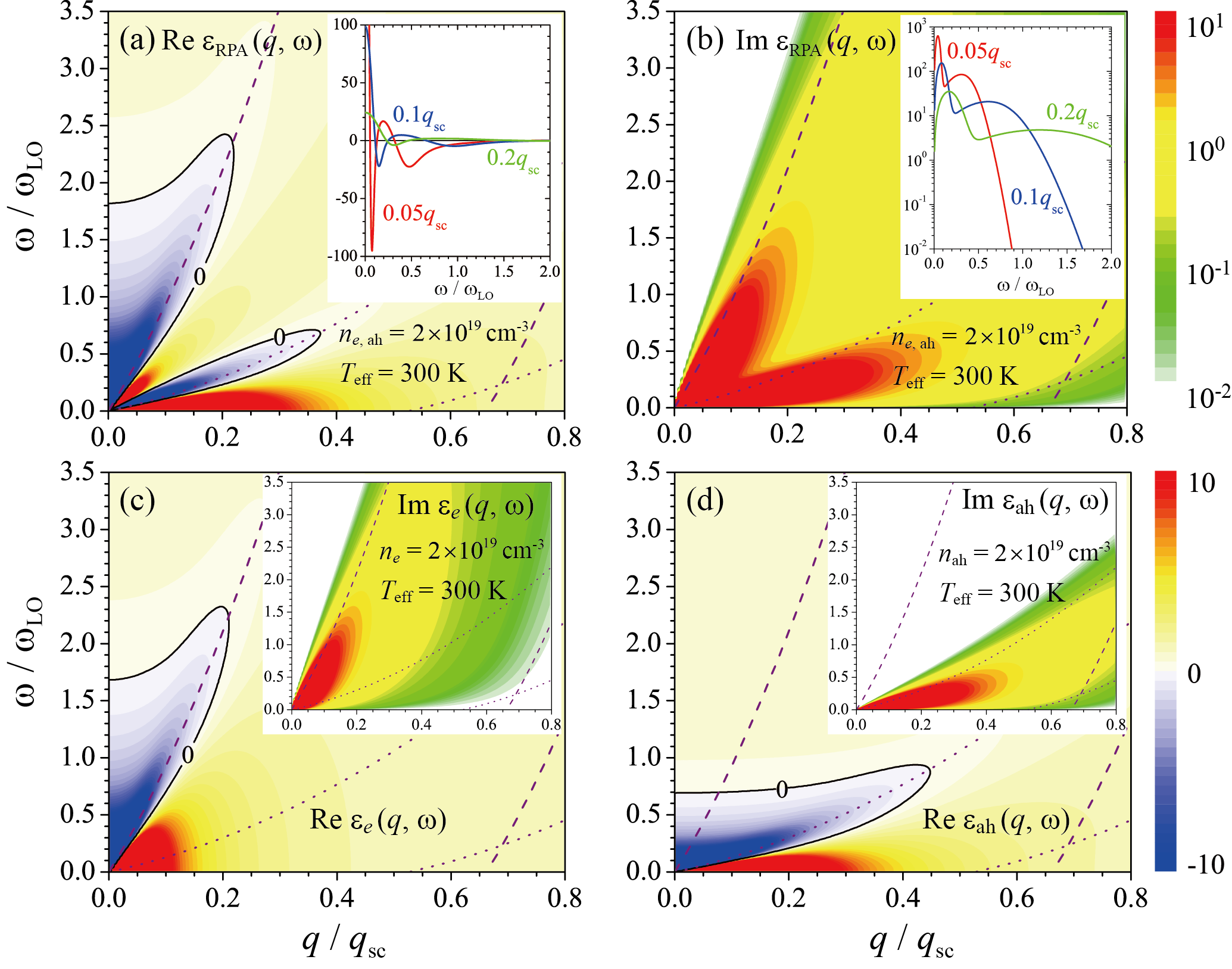}
\caption{Dispersive behavior of effective dielectric functions $\epsilon_{\rm RPA}(q,\omega)$ of a wurtzite GaN with conduction electron density of $2 \times 10^{19} \rm cm^{-3}$ at carrier effective temperature at 300 K: (a) $\mathcal{R}e ~\epsilon(q,\omega)$ and (b) $\mathcal{I}m ~\epsilon(q,\omega)$ for a multi-component plasma. Insets illustrate the frequency dependence of $\epsilon_{\rm RPA}(q,\omega)$ for representative wave numbers. (c) $\epsilon_{\rm e}(q,\omega)$ and (d) $\epsilon_{ah}(q,\omega)$ for a single-component plasma consisting of either electrons or A-holes, respectively. Insets show $\mathcal{I}m ~\epsilon_{e(ah)}(q,\omega)$, respectively. 
Pairs of dashed lines denote the corresponding regions of allowed single particle excitation continuum for electrons and A-holes in a wurtzite GaN.}
\label{epsilon-rpa}
\end{figure*}  

In a multi-component plasma in semiconductor materials, the dynamic screening of carriers introduces totally modified behavior in the dielectric reponse functions of the plasma through dielectric polarizations. 
At the carrier densities of $n_{\rm e} < n_{\rm th}$ and, hence, the B-hole band being empty in a wurtzite GaN, we assume that the photo-generated electrons and holes occupy the conduction and the A-hole valence bands with effective mass $m_{\rm e}=0.22 m_0$ and $m_{\rm A}=1.3 m_0$, respectively.  
Since the ratio of effective masses for the electrons and the holes are much different from the unity, the present electron-hole plasma can be treated as a two-component plasma.
In the case of two-component plasma of different effective masses, we expect the presence of multiple branches of modes, unlike the  single branch typically observed in the single-component plasma of doped semiconductors,\cite{Abstreiter1984}.
The spectral behavior of the carrier polarization functions $\Pi_c (q,\omega)$, in the presence of dynamic carrier screening, are illustrated in Fig. \ref{PI_ri_wq_300K} at $T_{\rm eff}= 300 \rm K$ for $n_{\rm e}=2 \times 10^{19} \rm cm^{-3}$ in a wurtzite GaN.
Pairs of dashed and dotted lines denote the allowed regions of single particle excitations for electrons and A-holes, respectively.
The spectral behavior is drastically modified from that of noninteracting case shown in Fig. \ref{P0_ri_wq_300K} or the case of dressed single-component plasma shown in Fig. \ref{PI_scp_ri_wq_300K}.
In addition to the contribution from the conventional optic plasmon-LO phonon coupling of the single-component plasma such as in doped semiconductors,\cite{DasSarma1990} the contribution from the low frequency acoustic collective oscillations occurs,\cite{Platzman1973} as is illustrated in the panels (a) and (c) of Fig. \ref{PI_ri_wq_300K} for $\mathcal{R}e \Pi_{e}(q,\omega)$ and $\mathcal{I}m \Pi_{e}(q,\omega)$.
The effects of the high frequency optic modes are clearly seen in $\mathcal{R}e \Pi_{e}(q,\omega)$ and $\mathcal{I}m \Pi_{e}(q,\omega)$ occurring at frequencies higher than that of the bare LO phonons well outside the single particle excitation continuum.
The low frequency acoustic plasmon branch is more broadened compared to the optic one, but is located outside the continuum of single particle excitations for A-holes.%[see panel (a).] 
The contribution of the heavier mass species (A-holes in the present work) to dressed $\Pi_c (q,\omega)$ [panels (b) and (d) of Fig. \ref{PI_ri_wq_300K}] is very similar to the case of bare polarization functions $\Pi_c^0 (q,\omega)$ represented by Eq.(\ref{Pi0}) and illustrated in Fig. \ref{P0_ri_wq_300K}.
The spectral behavior of  $\Pi_c (q,\omega)$ illustrated in Fig. \ref{PI_ri_wq_300K} reveal the multi-component band character of the high frequency optic and low frequency acoustic branches.  
Boundaries of a pair of single particle continua for electrons and A-holes are indicated with steeper dashed (for electrons) and slower dotted (for A-holes) lines, respectively, and branches of the optical and acoustical plasmon excitations are clearly distinguished in strong color intensities.
The mode of acoustic collective oscillation is expected to be Landau damped due to single particle excitations of the faster electrons, since they are located within the single particle excitation continuum of the lighter carriers. [See Fig.\ref{epsilon-rpa}(a).]   
Insets illustrate cross-sectional view of the spectral behavior at $q=0.1$ and 0.19 $q_{\rm sc}$, where $q_{\rm sc} = 2.58 \times 10^{7} \rm cm^{-1}$ is the Thomas-Fermi screening wave number at $T_{\rm eff}=300 \rm K$.
\begin{figure*}[t]
\includegraphics[width=14cm]{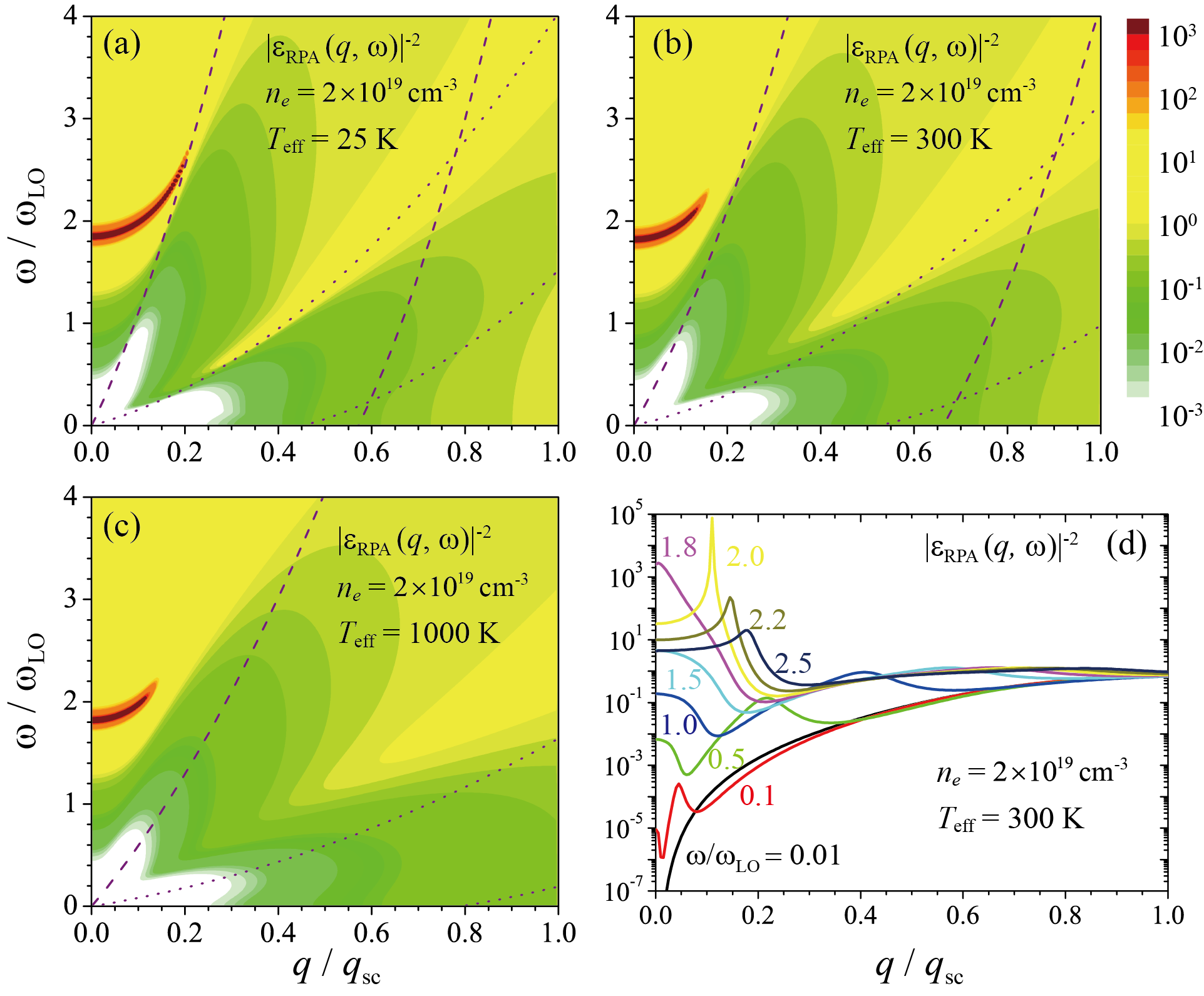}
\caption{Behavior of dressed dielectric response function $|\epsilon(q,\omega)|^{-2}$ of a wurtzite GaN with conduction electrons of $2 \times 10^{19} \rm cm^{-3}$ at carrier effective temperatures $T_{\rm eff}$ of (a) 25 K, (b) 300 K, (c) 1000 K, and (d) its dispersive behavior at 300 K for several inelastic frequency exchanges. Here $q_{\rm sc}$ is the Thomas-Fermi screening wave number at the corresponding temperature.
Pairs of dashed and dotted lines in (a) - (c) denote the corresponding regions of allowed single particle excitations for electrons and A-holes in a wurtzite GaN.}
\label{epsilon-2}
\end{figure*}

In Fig. \ref{epsilon-rpa}(a) and (b), dispersive behaviors of the real and imaginary part of effective dielectric function $\epsilon_{\rm RPA}(q,\omega)$ of Eq.(\ref{rpa epsilon}) for a wurtzite GaN with conduction electrons of $2 \times 10^{19} \rm cm^{-3}$ are shown for carrier effective temperature 300 K, respectively. Each inset illustrates the frequency dependence of $\epsilon_{\rm RPA}(q,\omega)$ for representative wave numbers.
The zero value contours of $\mathcal{R}e ~\epsilon(q,\omega)$ are indicated with dark solid lines, each denotes the dispersion curves of plasma oscillation modes.
Panels (c) and (d) show $\mathcal{R}e ~\epsilon^{\rm scp}(q,\omega)$ given by Eq.(\ref{epsilon single}) for the case of a single-component plasma consisting of either electrons or A-holes, respectively. Insets show $\mathcal{I}m ~\epsilon^{\rm scp}(q,\omega)$, respectively. 
Pairs of dashed and dotted lines denote the boundaries of allowed single particle excitation continua for electrons and A-holes in a wurtzite GaN.
In the region of high frequency and long wavelength, $\mathcal{I}m ~\Pi(q,\omega)$ vanishes and hence $\mathcal{I}m ~\epsilon(q,\omega)=0$, allowing self-sustaining collective oscillations with no dissipation.
In Fig. \ref{epsilon-rpa}(a), we find that a pair of collective modes are observed in the multi-component plasma, and that `optical' and `acoustic' plasma modes are well separated within the RPA, each damped through single particle excitations of electrons and holes, respectively. 
We note that, in panel (c), the bare carrier plasmon modes of frequencies $\Omega_{\rm e}\simeq 152 ~\rm meV =1.66 \omega_{\rm LO}$ and $\Omega_{\rm A} \simeq 46 ~\rm meV =0.5 \omega_{\rm LO}$ are well defined, but pushed apart from each other under the effect of dynamic screening in a multi-component plasma. [See Fig. \ref{epsilon-rpa}(a).]

In Fig. \ref{epsilon-2}(a)-(c), dispersive behaviors of the dressed dielectric response function $|\epsilon(q,\omega)|^{-2}$ of a wurtzite GaN with conduction electrons of $2 \times 10^{19} \rm cm^{-3}$ are shown for three different carrier effective temperatures 25 K (a), 300 K (b), and 1000 K (c), respectively.
Here the wave number is scaled by the Thomas-Fermi screening wave number $q_{\rm sc}$ at the corresponding temperature, $q_{\rm sc} = 1.73, 2.58, {\rm and}~ 2.97 \times 10^{7} \rm cm^{-1}$ at 1000, 300, and 25 K, respectively, each being $q_{\rm sc} \simeq 2.06, 3.07, {\rm and}~ 3.53 k_F$ with an electron concentration of $n_{\rm c} =2\times 10^{19} \rm cm^{-3}$. 
We observe that the effect of thermal broadening (temperature effect) on the response functions, especially on the damping of the collective modes, is much enhanced at higher temperatures.
Pairs of dashed and dotted lines denote the corresponding regions of allowed single particle excitation continuum for electrons and A-holes in a wurtzite GaN.
In Fig. \ref{epsilon-2} (d) we illustrate the nonlocal (finite $q$) behavior, i.e. cross sesctional view of $|\epsilon(q,\omega)|^{-2}$ at 300 K for several different values of inelastic frequency exchanges of $\omega=0.01 - 2.5 ~\omega_{\rm LO}$. 
For small frequency exchanges during carrier-carrier scattering, we note that the nonlocal, finite-$q$, effect of the screening is reduces to the case of Thomas-Fermi screening limit as is denoted by a monotonically increasing solid line for $\omega=0.01 ~\omega_{\rm LO}$ in panel (c).
        
\vspace{-0.5cm}
\section{Summary and Conclusion}
In this paper, the finite-temperature behavior of dielectric response functions of a multi-component hot carrier plasma is discussed.  
The effective dielectric function and polarizability functions are evaluated, employing random phase approximation, including the effects of dynamic screening at finite temperature. 
Spectral analyses of the carrier dielectric response functions are performed and their dynamic and nonlocal behaviors are studied. 
Our result, at high carrier densities such as of $n_{\rm e} \sim 2 \times 10^{19} \rm cm^{-3}$, shows that the well-defined optic plasmonic modes are clearly seen at frequencies higher than that of the LO phonons well outside the single particle excitation continuum and the additional low frequency acoustic plasmonic branch is located outside single particle excitations of heavier A-holes. 
The contribution of the heavier mass species to dressed $\Pi_c (q,\omega)$ is very similar to the case of bare polarization functions $\Pi_c^0 (q,\omega)$.
For small frequency exchanges during carrier-carrier scattering, the nonlocal, finite-$q$, effect of the screening is reduced to the case of Thomas-Fermi screening limit.

Finite-temperature behavior of multi-component solid-state plasma described here would be confirmed with such measurements of various photo-generated hot carrier spectroscopies as in wide band gap semiconductor materials.
The dielectric screening behavior introduced by carrier-carrier interaction would be utilized to investigation of the roles of various carrier-phonon coupling channels in the multi-component plasma material such as of polar and nonpolar optical phonons, acoustic deformation-potential and piezoelectrical Coulomb couplings.
Since the real part of the electric conductivity, a measure of dissipative processes, 
of the material is directly linked to the imaginary part of the retarded polarization function, 
our results can also be utilized to engineer the waveguide characteristics and plasmonics of the material.
The results we are presenting is suitable to experimental observation, and we hope that the present results could be verified, for example, by femtosecond optical measurements.

\begin{acknowledgments}
The authors acknowledge the support in part by Basic Science Research Program through the NRF of Korea  (grant number 201306330001). 
\end{acknowledgments}


\begin{thebibliography}{99}
\bibitem{Kyhm2011} K. Kyhm, L. Rota, and R.A. Taylor, Phys. Status Solidi A {\bf 208}, 1159 (2011).
\bibitem{Hwang2007} E. H. Hwang and S. Das Sarma, Phys. Rev. B {\bf 75}, 205418 (2007).
\bibitem{Wunsch} B. Wunsch, T. Stauber, F. Sols, and F. Guinea, New J. Phys. {\bf 8},318 (2006).
\bibitem{Prunnila2007} M. Prunnila, Phys. Rev. B {\bf 75}, 165322 (2007).
\bibitem{DasSarma1990} S. Das Sarma, J.K. Jain, and R. Jalabert, Phys. Rev. B {\bf 41}, 3561 (1990).
\bibitem{Platzman1973} P.M. Platzman and P.A. Wolff, {\it Waves and Interactions in Solid State Plasmas} (Academic Press, New York, 1973).

\bibitem{Abstreiter1984} G. Abstreiter, M. Cardona, and A. Pinczuk, {\it Light Scattering in Solids IV} ed. by M. Cardona and G. G\"untherodt, (Springer-Verlag, New York, 1984), and the references there in.

\bibitem{Fetter} A. Fetter and J.D. Walecka, {\it Quantum Theory of Many-Particle Systems}, (McGraw-Hill, 
New York, 1971).

\bibitem{Bruus}  H. Bruus and K. Fensberg, {\it Many-Body Quantum Theory in Condensed Matter Physics}, (Oxford Univ. Press, New York, 2004).
\bibitem{Mahan} See, for example, G. Mahan, {\it Many Particle Physics}, 3rd ed. (Plenum, 
New York, 2000).

\bibitem{Giuliani} G. Giuliani and G. Vignale, {\it Quantum Theory of Electron Liquid}, (Cambridge Univ. Press, 
New York, 2005).

\bibitem{Yi-Quinn1996} K.S. Yi and J.J. Quinn, Phys. Rev. B {\bf 54}, 13398 (1996).

\bibitem{Collet1993} J.H. Collet, Phys. Rev. B {\bf 47}, 10279 (1993).

\bibitem{Maldague} P.F. Maldague, Surf. Science {\bf 73}, 296 (1978).

\bibitem{De2010} A. De and C.E. Pryor, Phys. Rev. B {\bf 81}, 155210 (2010).

\bibitem{suzuki1995} M. Suzuki, T. Uenoyama, A. Yanase, Phys. Rev. B {\bf 52}, 8132 (1995).

\bibitem{chuang1996} S.L. Chuanga and C.S. Chang, Appl. Phys. Lett. {\bf 68}, 1657 (1996).

\bibitem{Collet1989} J.H. Collet, Phys. Rev. B {\bf 39}, 7659 (1989).

\bibitem{Landolt} Landolt-B\"ornstein: {\it Numerical Data and Functional Relationships in Science and Technology}, Vol. 17 Semiconductors, ed. by O. Madelung, M. Schulz and H. Weiss, (Springer Verlag, Berlin, 1982).

\bibitem{Binet1999} F. Binet, J.Y. Duboz, J. Off, and F. Scholz, Phys. Rev. B {\bf 60}, 4715 (1999).

\end{thebibliography}
\end{document}